\documentclass[fleqn,usenatbib]{mnras}
\usepackage{newtxtext, newtxmath}
\usepackage[T1]{fontenc}
\usepackage{graphicx}
\usepackage{amsmath}
\usepackage{makecell}
\usepackage{comment}

\title[]{Probing the Internal Structure of Neutron Stars: A Comparative Analysis of Three Different Classes of Equations of State}
\author[]{
Anshuman Verma,$^{1}$\thanks{E-mail:anshuman18@iiserb.ac.in} 
Asim Kumar Saha,$^{1}$\thanks{E-mail:asim21@iiserb.ac.in} Tuhin Malik,$^{2}$\thanks{E-mail:tuhin.malik@uc.pt} Ritam Mallick,$^{1}$\thanks{E-mail:mallick@iiserb.ac.in}
\\
$^{1}$ Indian Institute of Science Education and Research Bhopal, 462066, India\\
$^{2}$ Departamento de Física, Universidade de Coimbra, 3004-516 Coimbra, Portugal
}
\pubyear{\the\year{}}

\begin{document}
\label{firstpage}
\pagerange{\pageref{firstpage}--\pageref{lastpage}}
\maketitle

\begin{abstract}
Sound speed can be an important tool in unravelling the nature of matter that exists at the cores of neutron stars.
In this study, we investigate three major classes of equations of state; monotonous, non-monotonous and discontinuous depending on the nature of the sound speed in neutron stars. The monotonous EoS refers to hadronic models, the non-monotonous refers to the quarkyonic or smooth crossover models and discontinuous refers to discontinuous first-order phase transition models. We generate a large ensemble of EoS for three classes with the model agnostic speed of sound interpolation approach. Our main aim is to check which class of EoS is most favoured by present astrophysical bounds. It is seen that although non-monotonous and discontinuous is favoured thermodynamically, the usual neutron star observations like mass-radius, and f-mode oscillation fail to provide a satisfactory result. The universal relations are also seen to be futile as they show considerable spread and significant overlaps among the different classes. The Bayesian analysis shows slight bias towards the non-monotonous model but fails to provide a decisive answer.
\end{abstract}


\maketitle

\section{Introduction}
The study of neutron stars offers unique insights into the properties of dense matter, combining aspects of nuclear physics, astrophysics, and gravitational wave astronomy to explore matter under extreme conditions \citep{colaiuda_2008}. Neutron stars, compact remnants of supernovae with densities exceeding that of atomic nuclei, challenge our understanding of physics in high-density regimes \citep{Shapiro1983, Haensel2007}. Determining the equation of state (EoS) of neutron stars remains central to these studies, as the EoS dictates the relationship between pressure, density, and temperature, ultimately determining the observable mass and radius of these stars \citep{Oppenheimer1939, Tolman1939, Lattimer2001}.

Despite advancements, several questions persist regarding the nature of matter at densities well above nuclear saturation. Observational data, such as the 2.0 solar mass constraint set by PSR J0348+0432 \citep{Antoniadis2013,Cromartie2020} and the gravitational wave data from the GW170817 neutron star merger \citep{Rezzolla2016, Abbott2017}, provide essential constraints on viable EoS models \citep{Ozel2016, Abbott_2016, Riley:2019yda, Riley:2021pdl, Miller:2019cac, Miller:2021qha, Choudhury:2024xbk}. However, the possibility of quark matter phases or abrupt phase transitions remains unresolved \citep{Prakash2021}. Although they have been successful in constraining the EoS to a certain extent, fundamental problems still persist. It is not entirely clear whether the core of NSs harbours pure hadronic matter or there exist the possibility of exotic matter to be present. 

The primary objective of this paper is to determine which class of equation of state best describes the interior composition of neutron stars based on current astrophysical constraints. We investigate three major classes of EoS characterized by distinct speeds of sound behaviours: (1) monotonous EoS, representing traditional hadronic models with smoothly varying sound speed; (2) non-monotonous EoS, representing quarkyonic matter or smooth crossover models with peaked sound speed profiles; and (3) discontinuous EoS, representing first-order phase transitions with discontinuities in the sound speed. The significance of this work lies in our novel approach: rather than comparing specific microscopic models, we employ a model-agnostic speed of sound interpolation technique to generate large ensembles of EoS for each class \citep{Annala2020, Kojo2021,Chatterjee2024,verma2025}. This allows us to systematically evaluate which broad category of matter composition is most consistent with observations, moving beyond the limitations of specific model assumptions.


Hadronic EoS models are traditionally used to describe purely nucleonic matter, yielding predictions consistent with softer limits on neutron star stiffness at high densities \citep{Akmal1998, Gandolfi2014}. Such models, however, do not account for the possible deconfinement of quarks or other exotic states that may emerge at densities greater than a few times nuclear saturation \citep{Baym2018, Tews2018}. As an alternative, Quarkyonic EoS models introduce a phase where quark and nucleon matter coexist, representing a continuous phase transition that may better capture high-density behaviours \citep{McLerran2019, Jeong2020, Zhao2020}. On the other hand, discontinious phase transition EoS models, which incorporate first-order phase transitions, offer a more abrupt change between hadronic and quark phases, potentially corresponding to a more pronounced shift in neutron star properties \citep{Zdunik2013, Kurkela2010, Chatterjee2024}.

This study introduces a modified speed of sound interpolation method based on the framework developed by Annala et al. \citep{Annala2020, verma2025}, providing a unified approach for modelling EoS across a wide range of densities \citep{moriya_2016}. Unlike traditional polynomial or parametric models, the speed of sound interpolation allows for a consistent representation of the behaviour of high-density matter and preserves thermodynamic consistency across transitions. Moreover, the approach also allows for distinguishing EoS types by their speed of sound profiles \citep{Bedaque2015}, potentially indicating phase transitions through distinct peaks in sound speed at specified densities \citep{Alford2017, Annala2020}. Our approach spans a range of chemical potentials, aligning high-density predictions with perturbative quantum chromodynamics (pQCD) results at the highest relevant densities \citep{Freedman1977, Fraga2015}.

The implications of accurately modelling the EoS are important as observable properties of neutron stars, such as mass-radius relations, speed of sound, and oscillation frequencies, serve as potential indicators of the matter composition within their cores \citep{Lattimer2001, Steiner2010}. Recently, non-radial oscillation frequencies, especially the fundamental \( f \)-mode, have gained attention as they offer insight into the internal structure and stability of neutron stars \citep{Andersson1998, Kokkotas2001_a, Kokkotas2001_b,herbrik_2016, pratik2024}. By analyzing \( f \)-mode oscillations alongside mass-radius relationships and speed of sound profiles, this study aims to provide a comprehensive comparison of monotonous, non-monotonous, and discontinuous EoS models in the context of neutron star structure \citep{Issifu2023, Kokkotas2001_b, kantor_2014}.

This study aims to clarify the internal compositions and phase characteristics of neutron stars by comparing 10,000 sampled EoS from each model type and focusing on the feasibility of each EoS in matching astrophysical observations.  
With advancements in observational techniques, such as x-ray measurements of neutron star radii and continued gravitational wave observations, the findings of this research may assist in narrowing down the viable models for neutron star interiors, offering a benchmark for future theoretical and observational studies \citep{Glampedakis2018, Abbott2020, Raaijmakers_2019}.

We begin by constructing the macroscopic equations of state (EoSs) that will be used throughout this study. The structure of this paper is as follows: Section \ref{Formalism} discusses the EoSs classification, a brief overview of non-radial oscillations and the bayesian technique used to classify the EoSs. Sections \ref{Results_and_Discussion} focus on discussing the results. Finally, Section \ref{Summary_and_Conclusion} provides a summary and conclusion of the findings. For this study, we use natural units where \(G = 1\), \(c = 1\), and \(\hbar = 1\).

\section{Formalism}\label{Formalism}
\subsection{Construction of EoS}
The properties of matter below the nuclear saturation density ($n_0$) have been well studied. Hence, till densities of $\sim 0.5 n_0$, we use the tabulated BPS EOS as given by \citep{BPS}. Continuing till density of $\sim 1.1 n_0$ we use polytropes of the form $P = Kn^{\Gamma}$, where the value of $\Gamma \in [1.77, 3.23]$ to span the CET band simultaneously respecting the soft and stiff limits of \citep{Hebeler_2013}. Beyond this density, we adopt the speed of sound interpolation technique introduced by \citep{Annala2020}. This method parametrizes the speed of sound ($c_s^2$) as a function of chemical potential ($\mu$), which allows us to write the number density as 
\begin{equation}
n(\mu) = n_{\scriptscriptstyle CET}\exp\left[\int_{\mu_{CET}}^{\mu} \frac{d\mu'}{\mu' c_s^2(\mu')}\right]
\label{1}
\end{equation}
where $n_{CET}$ is fixed by matching the lower-density CET EOS to the initials of the interpolated EOS, and $\mu_{CET}$ is the chemical potential corresponding to $n_{CET}$. Pressure can then readily be calculated as
\begin{equation}
p(\mu) = p_{\scriptscriptstyle CET} \! + n_{\scriptscriptstyle CET}\int_{\mu_{CET}}^{\mu}d\mu' \exp\left[\int_{\mu_{CET}}^{\mu'}\frac{d\mu''}{\mu''c_s^2(\mu'')}\right]
\label{2}
\end{equation}   
where $p_{CET}$ corresponds to the pressure value at the matching point of the CET EOS. Our study here includes 5 randomised segments of ($c_{s,i}^2,\mu_{i}$), where $\mu_{i} \in [\mu_{\scriptscriptstyle CET},2.6$ 
 $GeV]$. The upper limit of $\mu$ is chosen by \citep{Kurkela_2014}, such that the uncertainty in the pQCD regime is roughly the same as the uncertainty at CET. A piecewise linear function is employed to connect the points $\left\{\mu_{i},c_{s,i}^{2}\right\}$ as:
\begin{equation}
c_s^2(\mu) = \frac{(\mu_{i+1} - \mu)c_{s,i}^2 + (\mu - \mu_{i})c_{s,i+1}^2}{\mu_{i+1} - \mu_i}
\label{3}
\end{equation}
where $c_{s,i}^2 \in [0,1]$. This function aids us to carry out the integrals of \eqref{1} and \eqref{2}. Our EOS used here does not continue to span pQCD limits as the densities where such phase is believed to exist are well beyond those, realised even at the cores of massive neutron stars.

To construct the discontinous (basically first-order phase transition (FOPT)) EOS, we modify Annala’s construction \citep{Annala2020} to suit our study better. While Annala's construction method inherently permits first-order phase transitions, they are statistically suppressed. The construction formalism laid out by \citep{verma2025}, allows for a more flexible representation of FOPT EoS. To address this, we modify the construction by dividing it into two segments: the first and second branches. The first branch, which mimics the hadronic branch is extended up to a chemical potential of $\mu_{\scriptscriptstyle PT}$; where $\mu_{\scriptscriptstyle PT}$ is chosen randomly in the range between $\mu_{\scriptscriptstyle CET}$ and onset of pQCD potential. 
The integrals of equation \eqref{1} and \eqref{2} for the second branch are then carried out from $\mu_{\scriptscriptstyle PT}$ to $\mu_{\scriptscriptstyle end}$ (2.6 GeV), where it is aligned with pQCD results \citep{Fraga_2014, Kurkela_2014}. Adhering to the astrophysical constraints of \citep{Demorest2010, John_Antoniadis}, we ensure that every EOS is capable of generating M$_{\scriptscriptstyle TOV}>2.0 M_\odot$.

\begin{figure}
    \centering
    \includegraphics[width=1.1\linewidth]{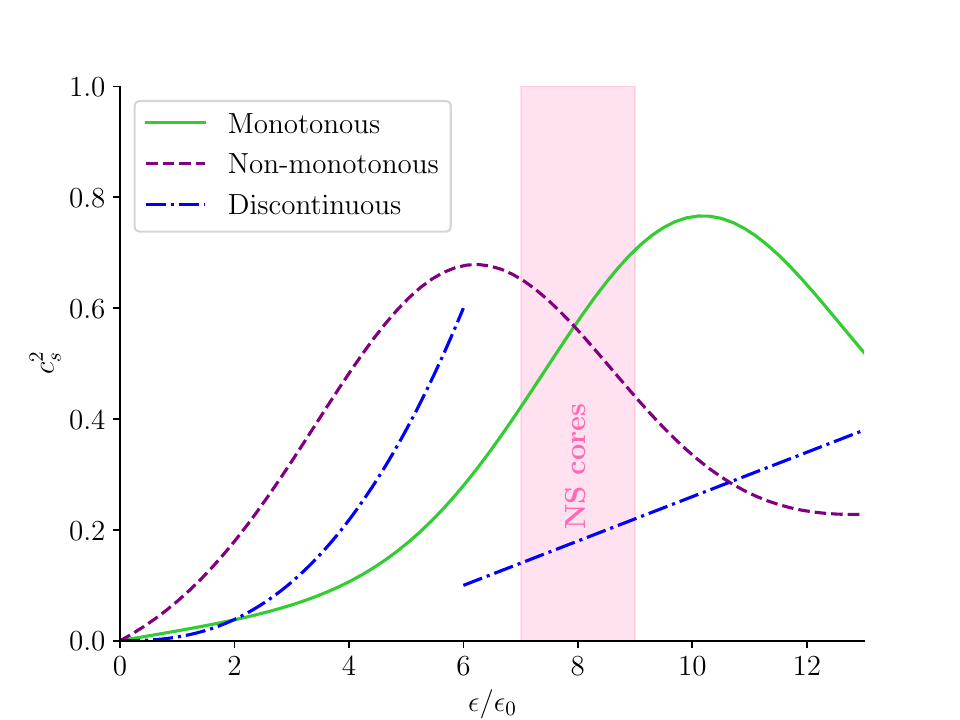}
    \caption{Schematic representation of the classification of EoS based on the presence of peaks in $c_s^2$.}
    \label{schematic}
\end{figure}

With an ensemble of EOS constructed, we then segregate the hadronic and quarkyonic EOS from our ensemble. This is done with the argument proposed by \cite{Annala:2023cwx}, which asserts $c_s^2$ as an indicator of phase transition. With \textcolor{black}{most} of our $M_{\scriptscriptstyle TOV}$ occurring in the density range of $\epsilon_{\scriptscriptstyle TOV} \sim \left(5\epsilon_0 - 8\epsilon_0\right)$, we apply the following segregation rule: EOS with a peak in $c_s^2$ that appears before or within this range are classified as \textit{non-monotonic} EoS, while those with the peak occurring after this range are identified as \textit{monotonic} EoSs. The discontinuous EoS are easily distinguishable from the other two as shown in Fig \ref{schematic}.
The monotonic EoSs comprise all hadronic EoS, while non-monotonic EoSs encompass quarkyonic, hyperonic or smooth crossover EoSs. Discontinuous EoS are those which have a discontinuity in the number density/energy density suitable for describing FOPT.

In addition to the observational bounds of $M_{TOV}>2M_\odot$, all the EOS models have been filtered out employing the tidal deformability constraint of GW170817\citep{Abbott2017}. It has been done by first evaluating the love number ($k_2$) of each stellar model along with their mass and radius as given by \citep{Hinderer_2010}. For an isolated star tidal deformability is given by $\Lambda = \frac{2}{3}k_2(\frac{R}{M})^5$. With $\Lambda$ calculated, binary tidal deformability ($\tilde{\Lambda}$) is then given by:\par
\begin{align}
    \tilde{\Lambda} = \frac{16}{13} \frac{(12M_2 + M_1)M_1{^4}\Lambda_1 + (12M_1 + M_2)M_2{^4}\Lambda_2}{(M_1 + M_2)^5},
\end{align}
where the subscripts 1,2 refer to the components of the binary system \citep{Leslie_Wade, Flanagan_2008}. As reported by the LIGO/Virgo data from the detection of GW170817 \citep{Abbott_2016}, $\tilde{\Lambda} < 720$ for $M_{chirp} = 1.186 M_{\odot}$ was employed to our EOS sets as well.

\subsection{The Non-Radial Oscillation}\label{Oscillation}

It has been seen that non-radial oscillation modes can also help in distinguishing EoS \citep{Kokkotas2001_a, Kumar:2021hzo,pratik2024}. In this study we check whether the f-modes can be used in distinguishing these three classes of EoS.
The non-radial oscillation analysis begins with the general metric for a spherically symmetric spacetime,

\begin{equation}
   ds^2 = e^{2\nu} dt^2 - e^{2\lambda} dr^2 - r^2 \left(d\theta^2 + \sin^2\theta \, d\phi^2\right),
\end{equation}

where \( \nu(r) \) and \( \lambda(r) \) are metric functions. The mass function \( m(r) \) is related to \( \lambda(r) \) through the equation,

\begin{equation}
   \lambda(r) =  - \frac{1}{2} \log\left(1 - \frac{2m}{r}\right).
\end{equation}

Using this metric, the structure of a spherically symmetric compact object can be derived by solving the Tolman–Oppenheimer–Volkoff (TOV) equations \citep{book.Glendenning1996},

\begin{align}\label{tov}
p' &=-(\epsilon + p) \frac{m + 4 \pi r^3 p}{r(r - 2m)}, \\
m' &= 4\pi r^2 \epsilon, \\
\nu' &= \frac{m + 4 \pi r^3 p}{r(r - 2m)},
\end{align}

where \( p(r) \) is the pressure, \( \epsilon(r) \) is the energy density, and the prime denotes the derivative with respect to \( r \). These equations are solved from the centre of the star (\( r = 0 \)) to its surface (\( r = R \)) with the following boundary conditions,
\begin{equation}\label{bc1}
   m(0) = 0, \quad p(0) = p_c, \quad p(R) = 0,
\end{equation}
and
\begin{equation}\label{bc2}
   e^{2\nu(R)} = 1 - \frac{2M}{R},
\end{equation}

where \( p_c \) is the central pressure, \( R \) is the radius where the pressure vanishes, and \( M = m(R) \) is the total mass of the star.

The star's oscillations can be explored by linearizing the relativistic Euler equation for small perturbations \citep{Kumar:2021hzo}. The resulting pulsation equations, which govern the oscillatory behaviour of the fluid inside the star, 
\begin{align}
Q' &= \frac{1}{c_s^2} \left[ \omega^2 r^2 e^{\lambda - 2\nu} Z + \nu' Q \right] - l(l+1) e^\lambda Z, \label{q}\\
Z' &= 2\nu' Z - e^\lambda \frac{Q}{r^2} + \frac{\omega_{\text{BV}}^2 e^{-2\nu}}{\nu'(1 - \frac{2m}{r})} \left( Z + \nu' e^{-\lambda + 2\nu} \frac{Q}{\omega^2 r^2} \right), \label{z}
\end{align}

where \( \omega_{\text{BV}}^2 \) is the Brunt-Väisälä frequency given by,
\begin{equation}
   \omega_{\text{BV}}^2 = {\nu'}^2 e^{2\nu} \left(1 - \frac{2m}{r}\right)\left(\frac{1}{c_e^2} - \frac{1}{c_s^2}\right),
\end{equation}

and \( c_e^2 = \left( \frac{dp}{d\epsilon} \right)_{y_i} \) and \( c_s^2 = \left( \frac{\partial p}{\partial \epsilon} \right)_s \) are the equilibrium and adiabatic sound speeds, respectively. The current study only explores the fundamental \( f \)-mode frequencies for all EoSs. Since the equilibrium and adiabatic sound speeds are identical for the \( f \)-modes, the Brunt-Väisälä frequency \( \omega_{\text{BV}} \) vanishes, making the last term in the equation for \( Z'(r) \) zero and eq.\ref{z} becomes, 
\begin{equation}
    Z' = 2\nu' Z - e^\lambda \frac{Q}{r^2} \label{zn}
\end{equation}
The above equations for the perturbing functions \( Q(r) \) and \( Z(r) \) are solved with appropriate boundary conditions, along with the TOV equations.

Near the center of the star, the behavior of \( Q(r) \) and \( Z(r) \) is given by,
\begin{equation}\label{bc3}
   Q(r) = Cr^{l+1}, \quad Z(r) = -\frac{C r^l}{l},
\end{equation}

where \( C \) is a constant, and \( l \) is the mode of the oscillation (for quadrupole modes, \( l = 2 \)). At the surface, the condition of vanishing Lagrangian pressure perturbation (\( \Delta p = 0 \)) leads to,
\begin{equation}\label{bc4}
   \omega^2 r^2 e^{\lambda - 2\nu} Z + \nu' Q \Big|_{r=R} = 0.
\end{equation}
By solving the coupled differential equations \eqref{q} and \eqref{zn}, along with the boundary conditions \eqref{bc1}, \eqref{bc2}, \eqref{bc3}, and \eqref{bc4}, we obtain the non-radial oscillation frequencies.

\subsection{Likelihood Estimation for Astrophysical Data}
However, the most robust check of the most favourable EoS in NS can only be done using statistical methods. This subsection outlines the Bayesian likelihood framework used to constrain the EoS of NS based on GW and X-ray observations. We detail the likelihood calculations for GW170817 and X-ray data from the Neutron Star Interior Composition Explorer (NICER), culminating in a combined likelihood for astrophysical constraints.

(i) {\it Gravitational Wave Observations: GW170817 - }
For GW observations, such as the binary neutron star merger GW170817, constraints on the EoS arise from the masses \( m_1 \) and \( m_2 \) of the binary components and their corresponding tidal deformabilities \( \Lambda_1 \) and \( \Lambda_2 \). The likelihood of the GW data given an EoS is expressed as:
\begin{align}
    P(d_{\mathrm{GW}}|\mathrm{EoS}) = \int_{m_2}^{M_u} dm_1 \int_{M_l}^{m_1} dm_2 \, P(m_1, m_2 | \mathrm{EoS}) \nonumber \\
    \times P(d_{\mathrm{GW}} | m_1, m_2, \Lambda_1 (m_1, \mathrm{EoS}), \Lambda_2 (m_2, \mathrm{EoS})) \nonumber \\
    = \mathcal{L}^{\rm GW},
    \label{eq:GW-evidence}
\end{align}
where \( P(d_{\mathrm{GW}} | m_1, m_2, \Lambda_1, \Lambda_2) \) represents the probability of the observed GW signal given the masses and tidal deformabilities, and \( \mathcal{L}^{\rm GW} \) is the resulting GW likelihood. The prior on the masses, \( P(m_1, m_2 | \mathrm{EoS}) \), is assumed to be uniform within physically motivated bounds (see, e.g., \cite{Agathos:2015uaa, Landry:2020vaw}). For a single mass \( m \), this prior is defined as:
\begin{equation}
    P(m | \mathrm{EoS}) = \begin{cases} 
        \frac{1}{M_u - M_l} & \text{if } M_l \leq m \leq M_u, \\ 
        0 & \text{otherwise},
    \end{cases}
\end{equation}
where \( M_l = 1 \, M_{\odot} \) and \( M_u = M_{\rm max} \), are the lower and upper mass bounds adopted in this analysis. 

(ii) {\it X-ray Observations: NICER -}
X-ray observations from NICER provide mass and radius measurements for neutron stars, offering complementary constraints on the EoS. The likelihood for X-ray data is given by:
\begin{align}
    P(d_{\rm X-ray} | \mathrm{EoS}) = \int_{M_l}^{M_u} dm \, P(m | \mathrm{EoS}) \nonumber \\
    \times P(d_{\rm X-ray} | m, R (m, \mathrm{EoS})) \nonumber \\
    = \mathcal{L}^{\rm NICER},
\end{align}
where \( P(d_{\rm X-ray} | m, R) \) is the probability of the observed X-ray data given a neutron star of mass \( m \) and radius \( R \), with \( R(m, \mathrm{EoS}) \) determined by the EoS. The mass prior \( P(m | \mathrm{EoS}) \) follows the same uniform distribution as above, but with bounds adjusted for NICER targets: \( M_l = 1.0 \, M_{\odot} \) (a typical lower limit for neutron stars) and \( M_u \) set to the maximum mass supported by the given EoS, which varies depending on the stiffness of the EoS.

In this work, we incorporate NICER observations of three pulsars: PSR J0030+0451 (NICER I), PSR J0740+6620 (NICER II), and PSR J0437+4715 (NICER III). Each contributes an independent likelihood term, \( \mathcal{L}^{\rm NICER \, I} \), \( \mathcal{L}^{\rm NICER \, II} \), and \( \mathcal{L}^{\rm NICER \, III} \), respectively, based on their measured mass-radius posteriors.

{\it Combined Astrophysical Likelihood - } The total likelihood combining GW and X-ray data is the product of the individual likelihoods, assuming independence of the observations:
\begin{equation}
    \mathcal{L} = \mathcal{L}^{\rm GW} \times \mathcal{L}^{\rm NICER \, I} \times \mathcal{L}^{\rm NICER \, II} \times \mathcal{L}^{\rm NICER \, III},
    \label{eq:finllhd}
\end{equation}
where \( \mathcal{L}^{\rm GW} \) encapsulates the GW170817 constraints, and the NICER terms reflect the mass-radius measurements from PSR J0030+0451, PSR J0740+6620, and PSR J0437+4715. This combined likelihood \( \mathcal{L} \) enables a robust Bayesian inference of the neutron star EoS by synthesizing multi-messenger astrophysical data.

\section{Results and Discussion}\label{Results_and_Discussion}

\begin{figure*}
    \centering
    \includegraphics[scale=0.50]{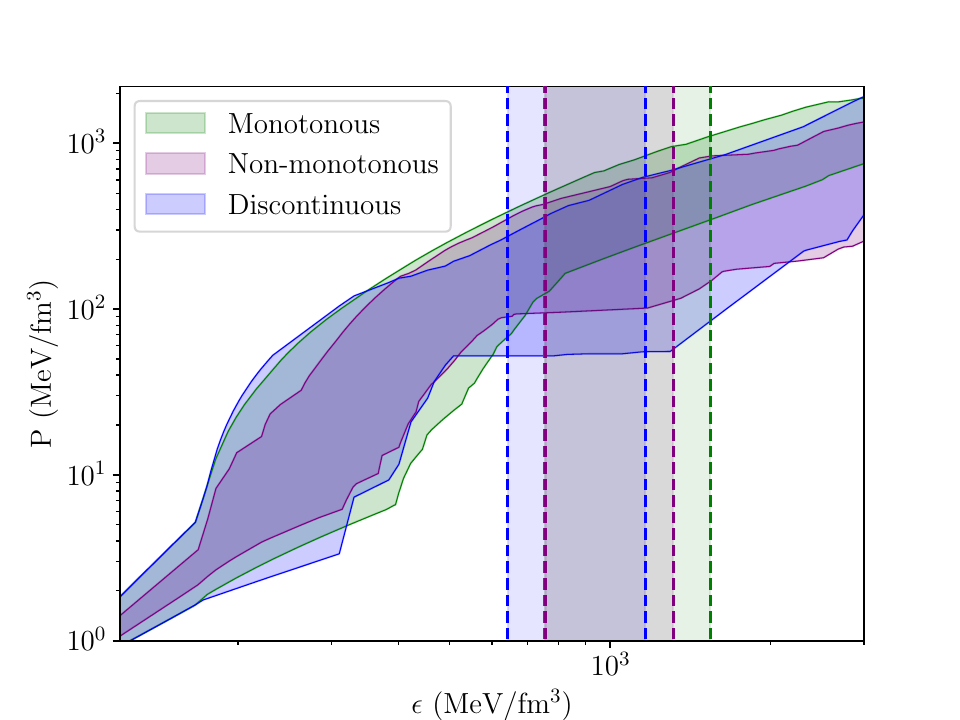}
    \includegraphics[scale=0.50]{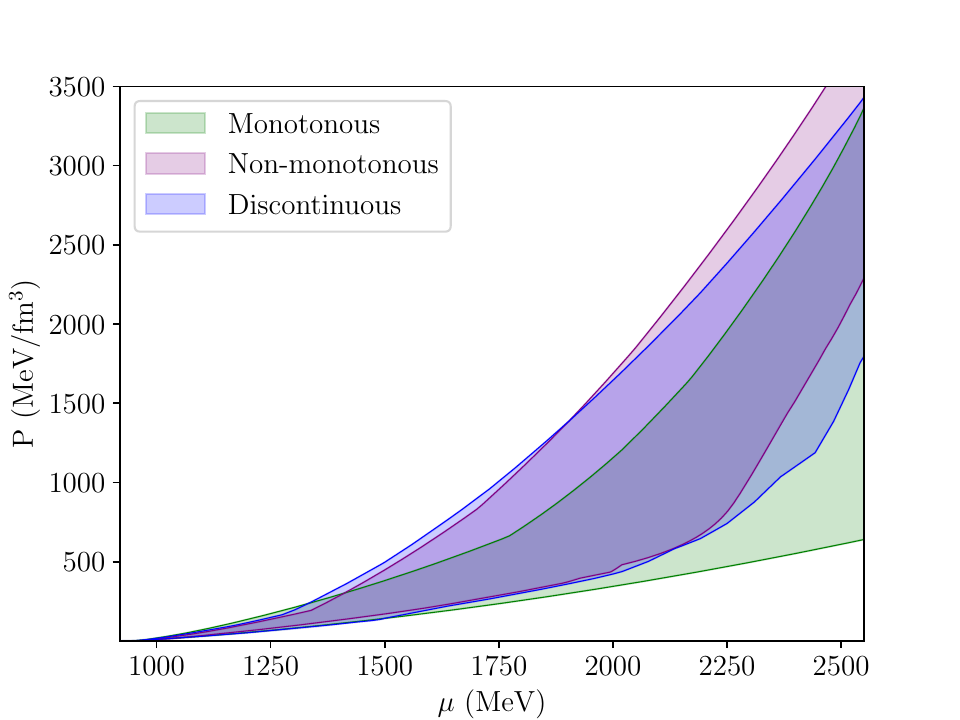}
    \caption{Figure: (Left) Pressure vs. Energy Density for different EoSs: The plot compares the behaviour of three EoS models — Monotonous (green), Non-monotonous (purple), and Discontinuous (blue) — using 10,000 sampled EoS for each set. Right: Pressure vs. Chemical Potential for different EoS are shown. The Monotonous class shows a gradual increase in pressure, while the Non-monotonous and Discontinuous EoS predict higher pressures at higher energy densities, representing the transition to quark matter. The overlapping regions represent uncertainties in the phase transition and quark matter regimes. The higher chemical potentials at a given pressure indicate thermodynamic stability. The Non-monotonous and Discontinuous class EoS exhibit higher chemical potentials at high pressures compared to the Monotonous, underscoring the stability of quark matter phases in dense neutron star cores. These results highlight the greater feasibility of hybrid EoS(Non-monotonous) models in describing the internal structure of neutron stars.}
    \label{eos_chem}
\end{figure*}

We performed a comparative analysis of three sets of EoS, 10,000 each - monotonic, non-monotonic and discontinuous to evaluate which model most feasibly describes the internal structure of neutron stars. The comparison is based on analyzing key properties such as mass-radius relationships, speed of sound behaviour, chemical potential distributions, and universal relations for each EoS. Finally, we do a Bayesian analysis to find the most likely EoS category given present astrophysical bounds. For the Bayesian analysis along with the maximum mass and tidal deformability bound we also use the present NICER constraints.

The left panel of Figure \ref{eos_chem} presents the pressure versus energy density relationship for the three sets of EoSs. The monotonous EoS (green band) demonstrates a gradual increase in pressure with rising energy density, representing the expected behaviour of purely hadronic matter at lower densities. The non-monotonous EoS (purple band), which may include quarks in a partially confined state, exhibits more complex behaviour at higher densities. The discontinuous FOPT EoS (blue band), shows a sharp rise in pressure at intermediate densities, signalling the phase change. While all models predict similar behaviour at lower energy densities following the chiral effective field theory (CET), the distinctions between them become more prominent at higher densities, highlighting uncertainties in the matter content at higher densities.

The relationship between pressure and chemical potential (Figure \ref{eos_chem}, right), the monotonous EoS remains softer at higher chemical potentials, suggesting that they are less favourable at high densities. The non-monotonous EoS, with its steeper rise in pressure, indicates greater thermodynamic stability at higher densities, while the discontinuous EoSs are more favourable than monotonous EoS but slightly less favoured than non-monotonous EoS at high densities.

\begin{figure*}
    \centering
    \includegraphics[scale=0.33]{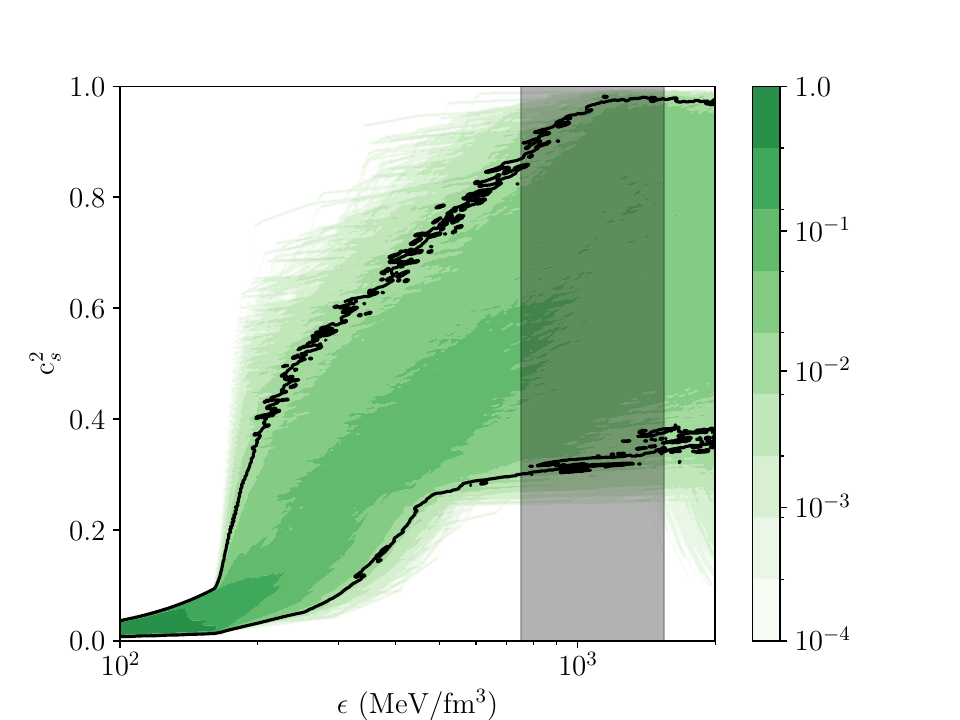}
    \includegraphics[scale=0.33]{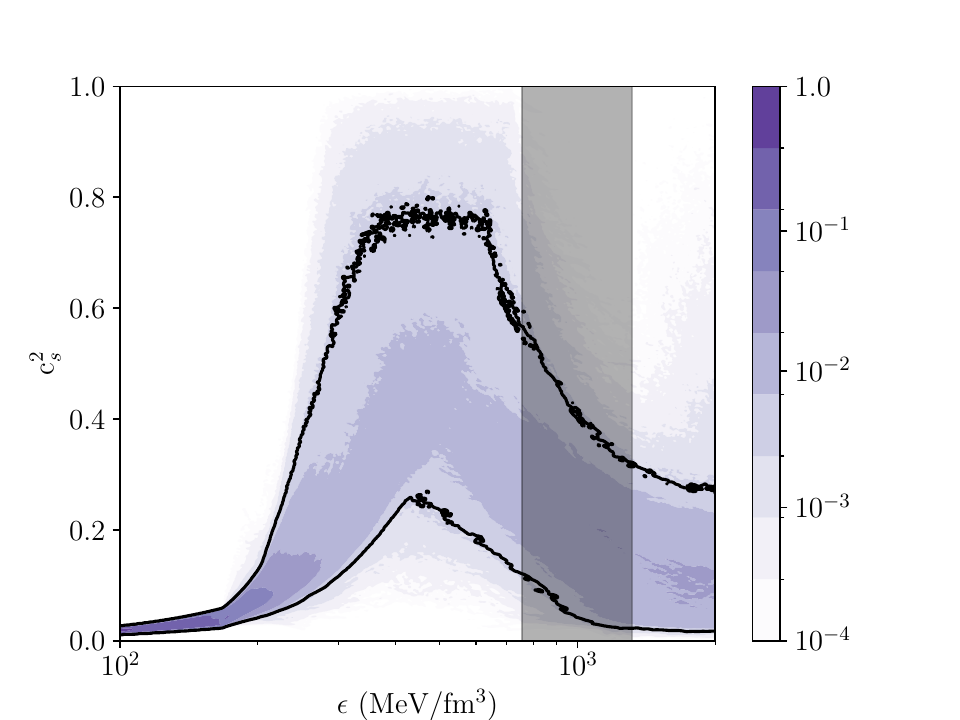}
    \includegraphics[scale=0.33]{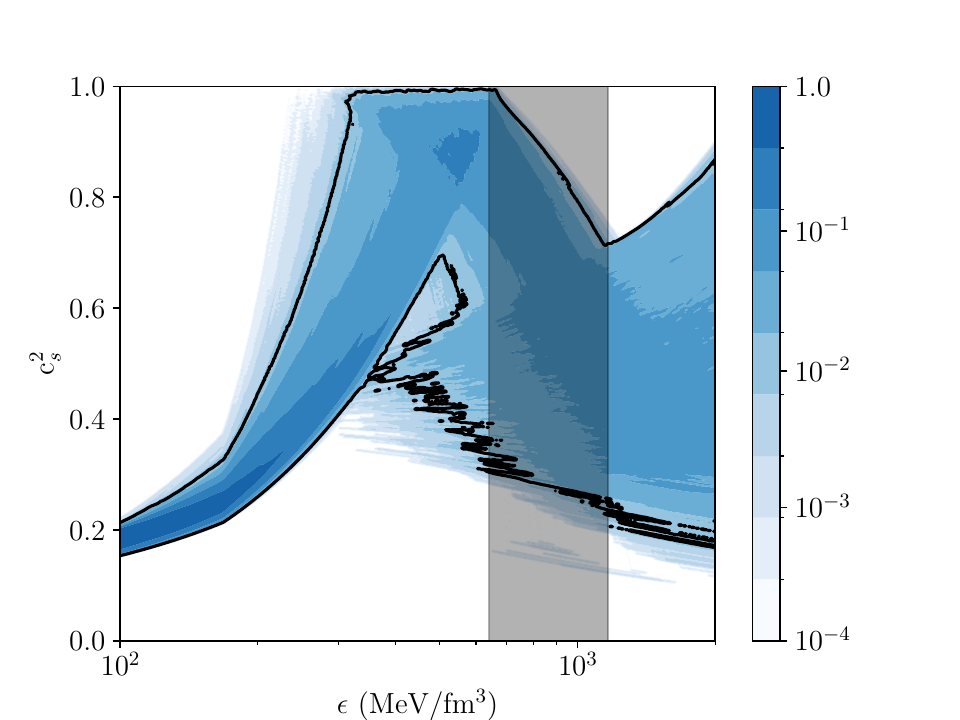}
    \caption{Probability density function (PDF) of the squared speed of sound, \( c_s^2 \), as a function of the energy density \( \epsilon \) for three different sets of equations of state (EOS). The grey translucent band highlights the range of densities corresponding to the central densities (\( \epsilon_{\text{TOV}} \)) of the maximum mass configurations. These plots illustrate the differences in stiffness and compressibility among the EoS models. (Left) Monotonous class EoS: shows a smooth and continuous increase in \(c_s^2\) with energy density, indicating a continuous stiffening of the EoS, with \(\epsilon_{\text{TOV}}\) in the range $\sim (5\epsilon_0 - 10\epsilon_0)$. (Middle) Non-monotonous class EoS: displays more complex behaviour, including a peak at intermediate densities, suggesting a crossover from hadronic to quark matter, with \(\epsilon_{\text{TOV}}\) in the range $\sim (5\epsilon_0 - 8.5\epsilon_0)$. (Right) Discontinuous class EoS: features a sharp drop in \(c_s^2\) at specific energy densities, consistent with a sudden phase transition, with \(\epsilon_{\text{TOV}}\) in the range $\sim (4\epsilon_0 - 7.7\epsilon_0)$. Across all panels, the \(3\sigma\) confidence region is outlined in black.}
    \label{eos_cs2}
\end{figure*}

The probability density functions (PDFs) of the speed of sound (\(c_s^2\)) provide a statistical representation of how different EoSs respond to variations in energy density, offering insights into their phase structure. Figure \ref{eos_cs2} presents these distributions for the three EoS classes. In the monotonous class (left panel), \(c_s^2\) increases smoothly with energy density, reflecting the absence of phase transitions and the continuous nature of hadronic matter. The non-monotonous class (middle panel) exhibits a peak at intermediate energy densities, indicative of a gradual transition from hadronic to quark matter, characteristic of a smooth crossover scenario. In contrast, the discontinuous class (right panel) shows a sharp rise followed by a sudden drop in \(c_s^2\), a signature of a first-order phase transition (FOPT).  
The non-monotonous class exhibits a broader peak, reflecting a more gradual transition between phases, distinguishing it from the discontinuous class, where the sharp peak signifies an abrupt FOPT.

\begin{figure*}
    \centering
    \includegraphics[scale=0.50]{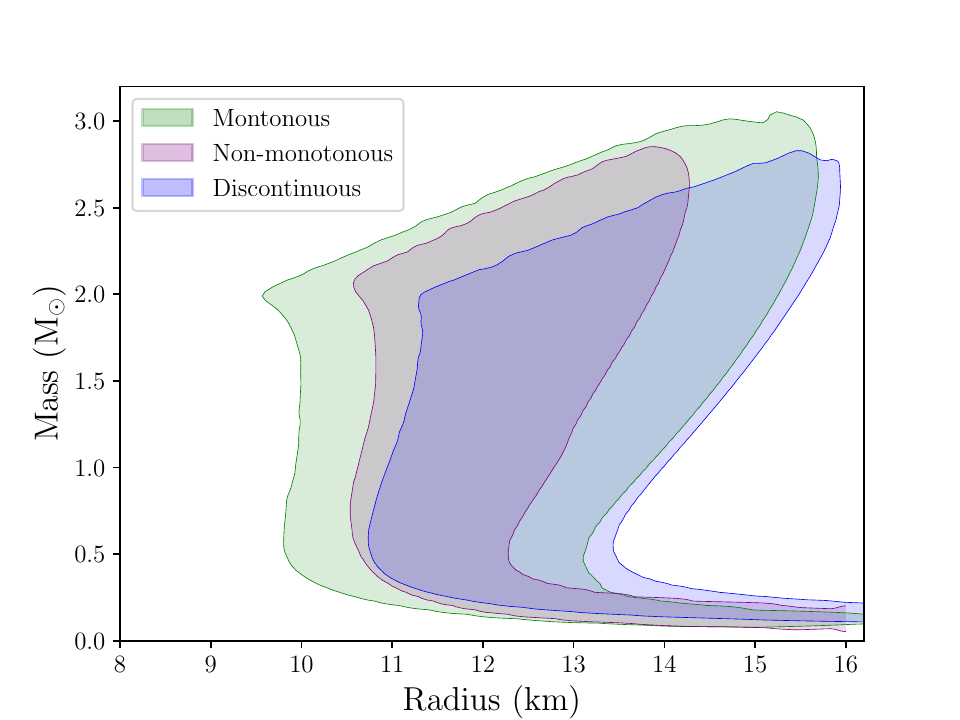}
    \includegraphics[scale=0.50]{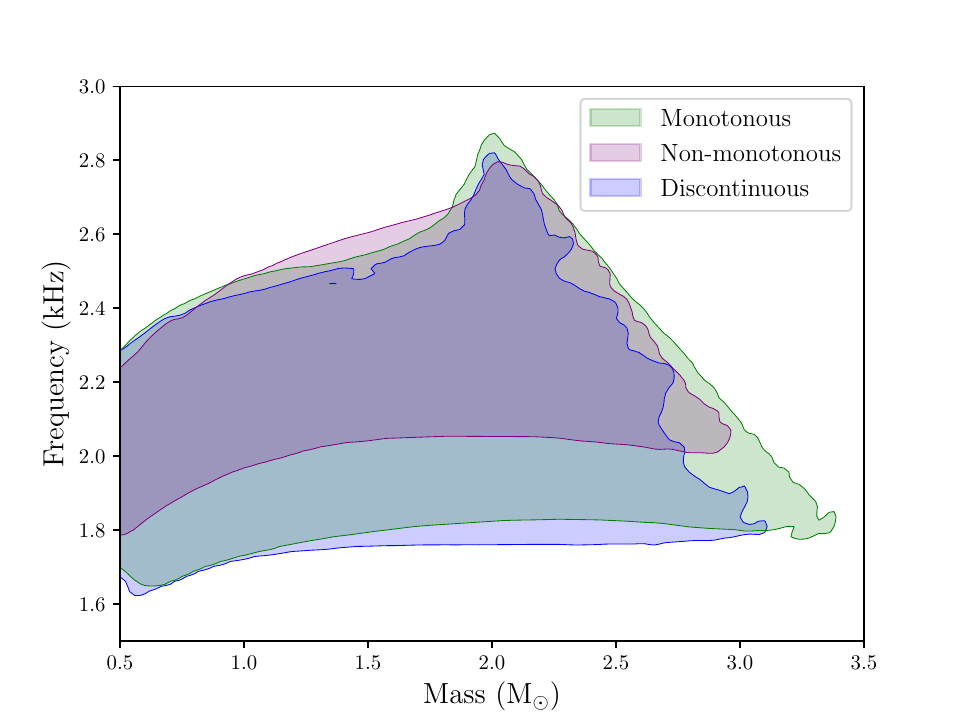}
    \caption{In this plot (Right panel): f-mode Oscillation Frequencies as a Function of Neutron Star Mass and, (Left panel) Mass-Radius (M-R) Relation for different EoSs are shown. The f-mode oscillation frequency versus mass for neutron stars with Monotonous (green), Non-monotonous (purple), and Discontinuous (blue) class EoSs. The non-monotonous and discontinuous models show broader, stable frequency ranges, especially at higher masses, indicating stronger binding and greater compactness, suggesting their thermodynamic feasibility over the Monotonous model. Similarly, the mass-radius curves derived from the same EoS sets demonstrate the effects of different internal compositions on neutron star properties. The monotonous class EoS spans from smaller to higher maximum masses and radii. At the same time, the non-monotonous and discontinuous class EoS allow for more massive neutron stars with larger radii, especially in the high-mass regime. The discontinuous class also suggests the possibility of even larger neutron stars, with larger radii.}
    \label{freq_mr}
\end{figure*}

The mass-radius (M-R) relationship (Figure \ref{freq_mr}, right) is crucial for understanding the observable properties of neutron stars. The monotonous EoS predicts a wide range of possible maximum masses, from 2.0 to 3.2 solar masses, with corresponding radii between 9 and 15 km.
In contrast, due to an additional degree of freedom, the non-monotonous EoS predicts more compact stars, with slightly smaller radii for a given mass, especially for more massive stars. The discontinuous EoS shows a broader range of radii, particularly for stars with masses above 2.0 solar masses, suggesting that a phase transition to quark matter could lead to larger neutron stars. 

The f-mode oscillation of NSs has been presently used to distinguish characteristically distinct EoS models \citep{Kokkotas2001_a, Kokkotas2001_b,fuller_2020, Kumar:2021hzo}. However, the distinction is not very prominent when one encompasses the whole ensemble of possible EoS.
Figure \ref{freq_mr} (left panel) provides the allowed band for the three sets of EoS. The monotonous EoS extends to all possible frequency ranges from low to high. It almost engulfs both the non-monotonous and discontinuous oscillation frequency ranges. The non-monotonous oscillation frequency lies mostly at the higher values; however, the discontinuous set oscillation frequency almost coincides with the monotonous band. 

Universal relation (UR) has emerged as an important tool to characterize different EoS. These relations are blind towards the microscopic structure of similar EoS types and give almost a UR among various dimensionless parameters of NS properties. However, they may differ for substantially different EoSs \citep{Zhao2020, Manoharan_2024}. These relations link parameters such as tidal deformability, f-mode oscillation frequency, average density, and compactness. By comparing these properties across various EoS classes, we can identify observational patterns and relationships that are unique to each type of EoS. This approach provides a valuable tool for understanding the internal structure of neutron stars based on measurable quantities.

\begin{figure*}
    \centering
    \includegraphics[scale=0.50]{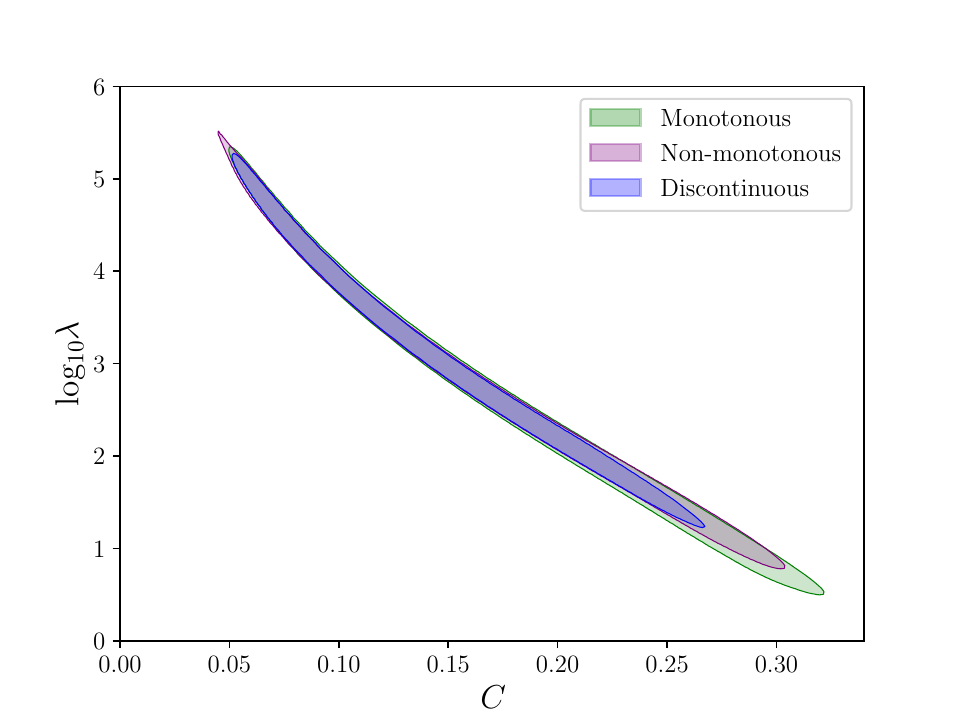}
    \includegraphics[scale=0.50]{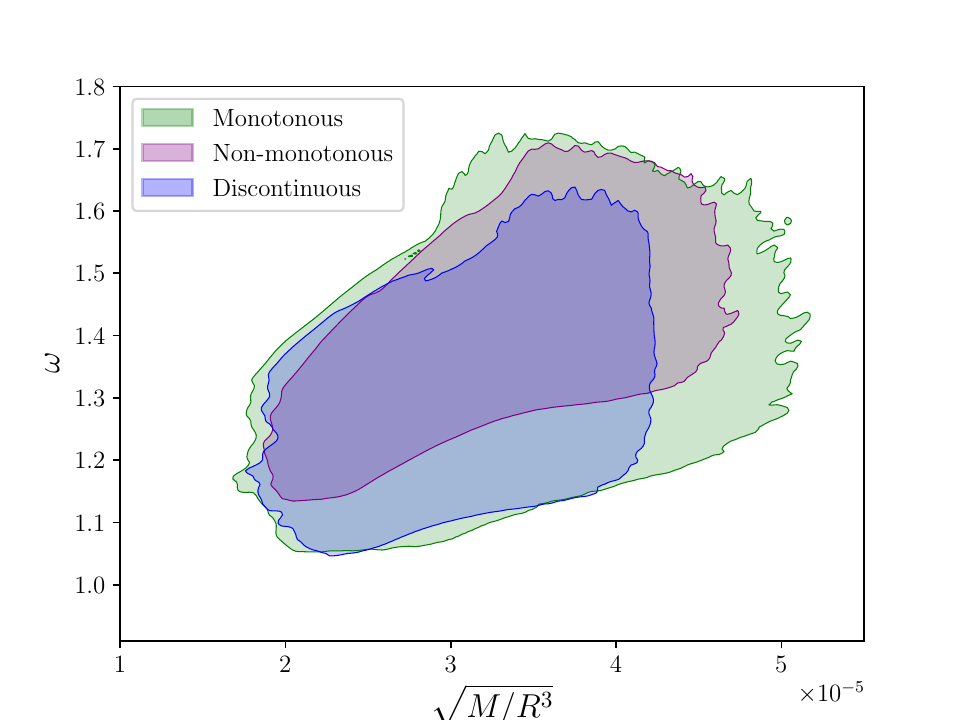}
    \includegraphics[scale=0.50]{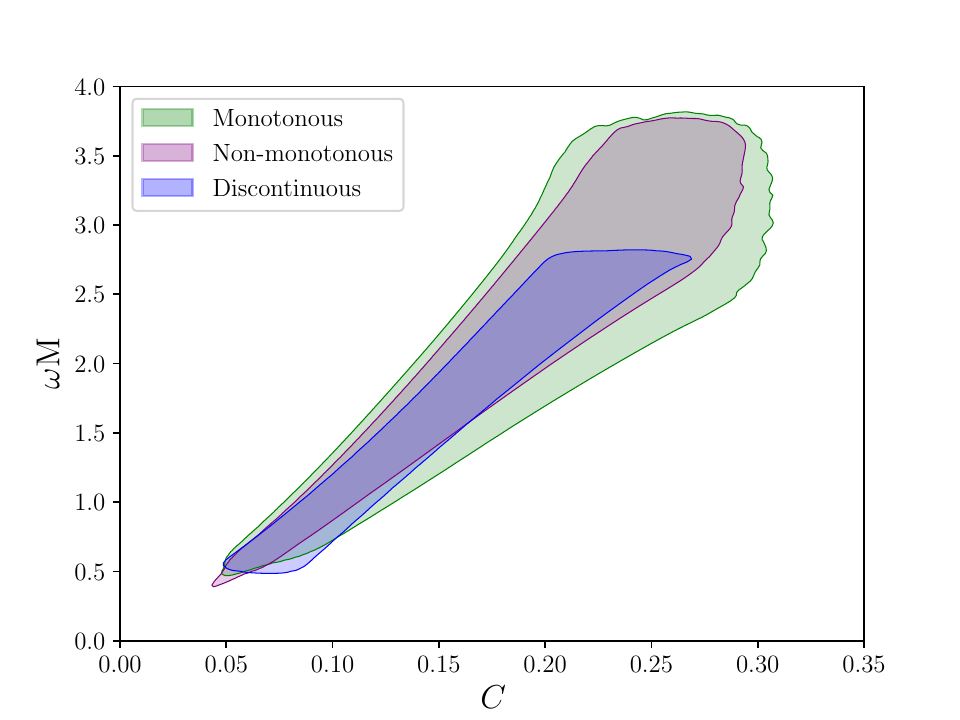}
    \includegraphics[scale=0.50]{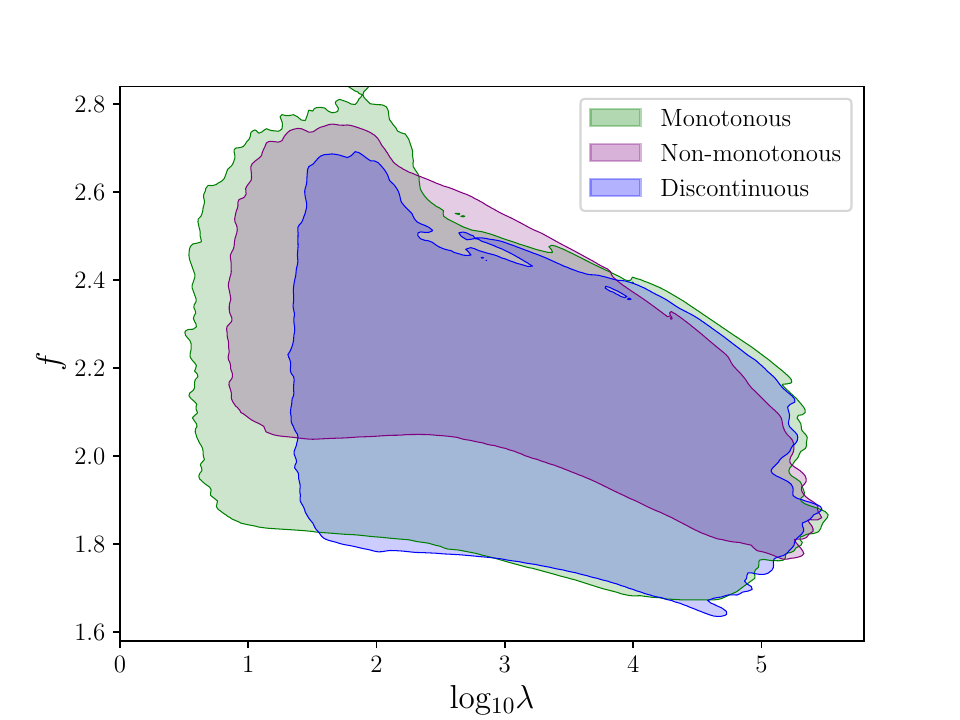}
    \caption{The upper-left panel depicts the relationship between the logarithm of tidal deformability (\(\log_{10} \lambda\)) and compactness (\(C = M/R\)), while the upper-right panel illustrates the correlation between angular frequency (\(\omega = 2\pi f\)) and the square root of the average density (\(\sqrt{M/R^3}\)). The lower-left panel presents the contour of the dimensionless parameter (\(\omega M\)) as a function of compactness (\(C\)), whereas the lower-right panel shows the variation of frequency (\(f\)) with \(\log_{10} \lambda\). Each panel compares results for three distinct classes of EoSs: Monotonous (green), Non-monotonous (purple), and Discontinuous (blue). The significant overlap observed in the upper-left panel suggests that the compactness–tidal deformability relation remains largely independent of the specific EoS, supporting its universal nature. However, in the other panels, a considerable spread in the data indicates that these relations are not strictly universal. While certain EoSs may individually follow a universal trend, this consistency diminishes when a broader set of EoSs is considered. Consequently, the substantial overlap among different EoS categories highlights the challenges in distinguishing between them using these universal relations alone.}
    \label{U1}
\end{figure*}

Figure \ref{U1} presents various universal relations (URs) associated with neutron star properties. The upper-left panel illustrates the relationship between compactness (\(C = M/R\)) and tidal deformability (\(\lambda\)) across three distinct classes of equations of state (EoSs). The significant overlap among the contours corresponding to different EoS classes suggests that this relation is largely independent of the specific microphysical details governing each EoS. 

The upper-right panel of Figure \ref{U1} depicts the correlation between the square root of the mean density (\(\sqrt{M/R^3}\)) and angular frequency (\(\omega\)) for various EoS classes. Unlike the compactness–tidal deformability relation, this UR exhibits considerable scatter, indicating that the universality is not as well-preserved. Although distinct EoS classes form somewhat separate regions, their substantial overlap suggests that astrophysical observations may not be sufficient to precisely distinguish between them.  

The lower-left panel of Figure \ref{U1} examines the relationship between compactness (\( C = M/R \)) and the dimensionless frequency parameter (\( \omega M \)) for the three EoS classes. Considering the whole ensemble of EoS the universality disappear. The overlapping nature of the contours further highlights the challenge in differentiating between the EoS classes based on this relation alone.  

Finally, the lower-right panel of Figure \ref{U1} presents the dependence of tidal deformability (\(\lambda\)) on frequency (\(f\)) across the three EoS categories. Similar to the previous cases, a significant spread is observed, accompanied by substantial overlap between the EoS classes. This widespread overlap further demonstrates the difficulty in distinguishing between different EoS models solely based on URs, even when considering various neutron star observables.

\begin{table*}
    \centering
    
    \begin{tabular}{c|c|c|c|c} 
        \hline
        EOS type & \makecell{GW} & \makecell{GW+\\J0030} & \makecell{GW+J0030+\\J0740} & \makecell{GW+J0030+J0740\\ + J0437}\\
        \hline
        Monotonous  & 9308 & 9242 & 8864 & 4225\\
        \hline
        Non-monotonous & 10000 & 9999 & 9998 & 7942 \\
        \hline
        Discontinuous   & 9987 & 9987 & 9957 & 5213 \\
        \hline
    \end{tabular}
    \caption{Statistics of the EOS ensemble for three sets, each adhering to different observational constraints.}
    \label{table stats}
\end{table*}

\subsection{Comparison of Equation of State Models Using Astrophysical Data}
As it is very difficult to differentiate three very different EoS classes from the usual NS observation, we finally resort to Bayesian analysis to distinguish them.

\begin{figure*}
    \centering
    \includegraphics[width=1\linewidth]{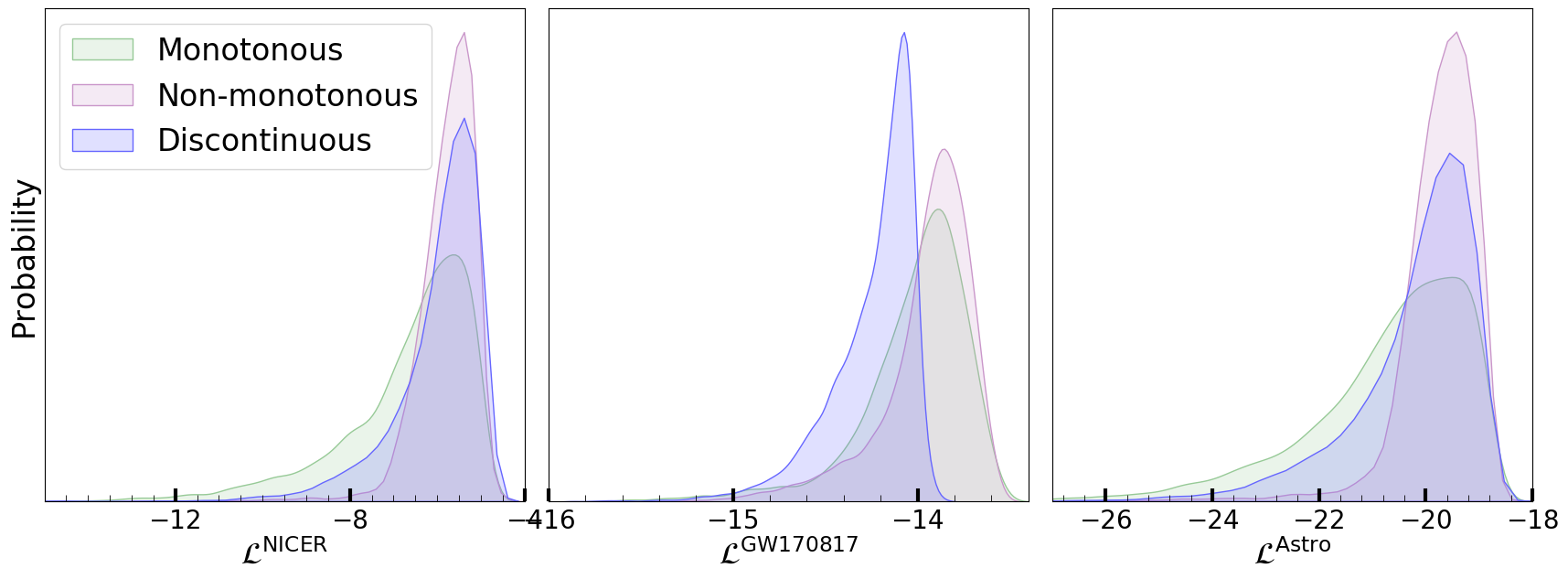}
   \caption{Log-likelihood distributions for the three sets considered in this work---Hadronic, Quarkyonic, and PT (phase transition)---for different astrophysical datasets. From left to right, the panels correspond to: (i) NICER observations (including PSR J0030+0451, PSR J0740+6620, and PSR J0437+4715); (ii) tidal deformability constraints from GW170817; and (iii) the combined astrophysical data (including both NICER and GW observations).}
    \label{fig:lnlike}
\end{figure*}

Figure \ref{fig:lnlike} illustrates the log-likelihood distributions for these three EoS models across three datasets: (i) NICER observations (combining mass-radius measurements for the three pulsars), (ii) GW170817 observations (focusing on tidal deformability constraints), and (iii) the combined astrophysical dataset (integrating both NICER and GW data). Each panel shows the log-likelihood (\( \mathcal{L}_{\text{NICER}} \), \( \mathcal{L}_{\text{GW170817}} \), and \( \mathcal{L}_{\text{Astro}} \)) for the monotonous (green), non-monotonous (pink), and discontinuous (blue) models. The distributions reveal the relative compatibility of each EoS with the observational data. For NICER data, the monotonous model shows a slightly broader distribution, while the non-monotonous and discontinuous models exhibit sharper peaks, indicating tighter constraints. For GW170817, the discontinuous model appears to dominate with the highest likelihood, suggesting it better accommodates the tidal deformability constraints. In the combined astrophysical likelihood, the ono-monotonous model emerges as the most favoured, with a peak likelihood surpassing the other two, as evident from the overlapping distributions.

To quantify the preference for each model, we compute the log evidence \( \ln(\mathcal{Z}) \) using Bayesian inference, with the previously generated EoS models serving as priors. Table \ref{tab:bayes_evidence} summarizes the log evidence values for the three models, providing a statistical measure of their overall fit to the combined astrophysical data. The non-monotonous model achieves the highest log evidence (\( \ln(\mathcal{Z}) = -19.56 \)), followed closely by the discontinuous model (\( \ln(\mathcal{Z}) = -19.80 \)), while the monotonous model has the lowest (\( \ln(\mathcal{Z}) = -20.03 \)). These results suggest that the non-monotonous EoS is the most consistent with the combined NICER and GW170817 data, though the differences are relatively small, indicating that all three models remain plausible within the current observational uncertainties.


\begin{table}
\setlength{\tabcolsep}{30.pt}
\renewcommand{\arraystretch}{1.2}
    \centering
    \caption{Log evidence $\ln (\mathcal{Z})$ Values for the different Models.}
\begin{tabular}{lr}
\hline \hline 
\textbf{Model}            & $\ln (\mathcal{Z})$       \\ \hline
Monotonous        &     -20.03    \\ 
Non-monotonous      &     -19.56    \\
Discontinuous              &     -19.80   \\  \hline
\end{tabular}
\label{tab:bayes_evidence}
\end{table}

\begin{figure}
    \centering
    \includegraphics[scale=0.45]{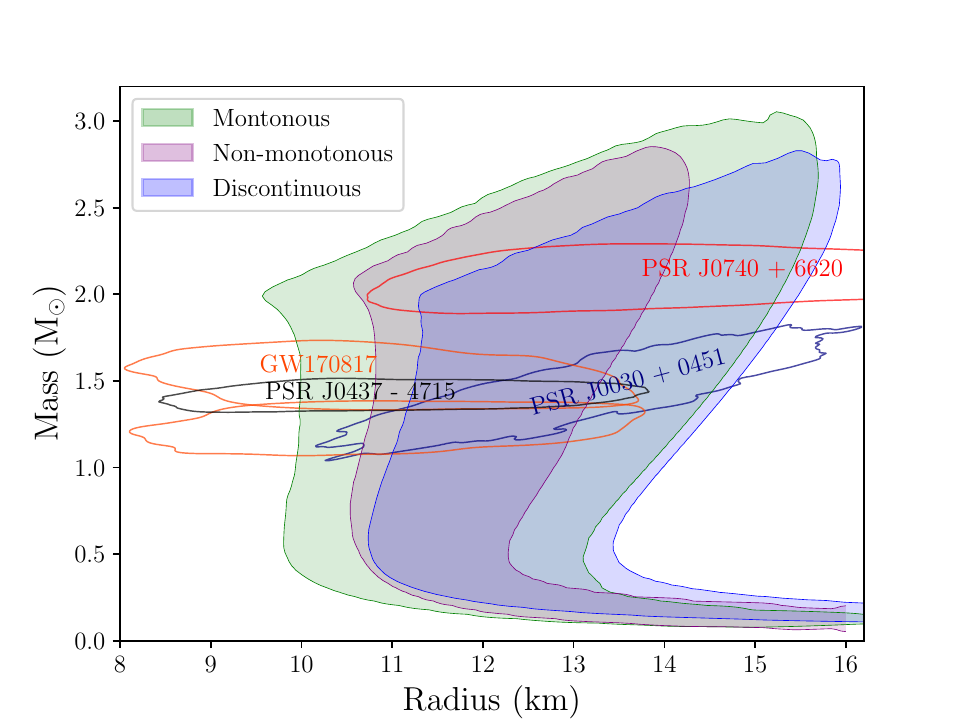}
    \includegraphics[scale=0.45]{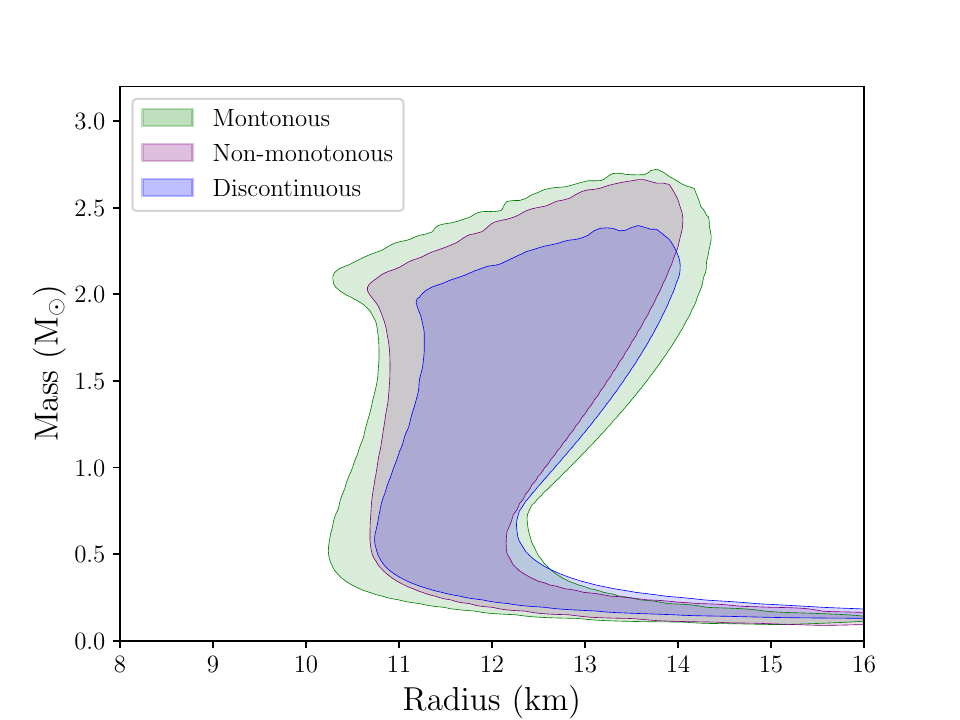}
    \caption{The MR region of three different ensembles of EoS along with the observational contours of GW170817(orange contour), PSR J0437+4715(black contour), PSR J0030+0451 (navy blue contour), PSR J0740+6620 (red contour). The lower panel compares the contours after imposing the observational bounds.}\label{observation}
\end{figure}

The results of Bayesian analysis can also be understood from the MR curve. Although the MR contour for three different EoS classes is substantially different (fig \ref{observation} (top panel)), the Bayesian analysis paints a different picture. This is because once we incorporate the NICER results, the GW170817 contour of the MR diagram of the three different classes almost overlaps (fig \ref{observation} (bottom panel)). Only through Bayesian analysis one then can have a result of the most probable EoS class, but even then all of them lie below the "substantial evidence" threshold \cite{jeffreys1998theory}. 

\section{Summary and Conclusion}\label{Summary_and_Conclusion}
This study classifies EoSs into three distinct categories based on the behaviour of the speed of sound in neutron star interior. The monotonous class, representing hadronic EoSs, exhibits a continuously increasing speed of sound. The non-monotonous class, which includes quarkyonic (smooth crossover) EoSs, features a peak in the speed of sound at intermediate densities before reaching the neutron star core. The discontinuous class, associated with FOPT EoSs, is characterized by an abrupt discontinuity in the speed of sound.  

An analysis of the EoS phase space indicates that the non-monotonous class appears more thermodynamically favourable at higher densities. The speed of sound contours reveals distinct behaviours across the three categories, with the non-monotonous and discontinuous classes exhibiting plateau-like regions. The MR relations and f-mode frequency versus mass distributions also show variations, with the monotonous and discontinuous classes showing the greatest spread. Although the non-monotonous class has the least spread, it still overlaps significantly with the other two.  

To distinguish between these EoS classes, various URs were examined. However, many showed considerable scatter and overlap, reducing their efficacy as diagnostic tools. The persistence of degeneracy across different URs highlights the challenge of uniquely identifying the internal composition of neutron stars using current observational constraints.  

Bayesian inference was employed to assess the most probable EoS class based on astrophysical data, including recent constraints from NICER, maximum mass limits, and tidal deformability measurements. Although the non-monotonous class demonstrated better agreement with these observations, the statistical evidence remained below the threshold for a definitive conclusion. This suggests that, at present, no single EoS class can be conclusively favoured over the others.  

Overall, despite clear theoretical distinctions among these EoS models, existing astrophysical observations do not provide sufficient evidence to differentiate them conclusively. While Bayesian analysis indicates a slight preference for the non-monotonous EoS, the statistical significance remains weak. This highlights the need for more precise observational constraints to resolve the degeneracy among EoS models. Future missions such as eXTP, New Athena, and the Einstein Telescope are expected to provide crucial insights through high-precision X-ray measurements of neutron star radii and masses, along with detailed gravitational wave observations from binary mergers. The classification framework presented in this study establishes a robust foundation for incorporating these upcoming observations, furthering our understanding of ultra-dense matter in neutron star interiors.

\section{Acknowledgments}
The authors thank the Indian Institute of Science Education and Research Bhopal for providing all the research and infrastructure facilities. A.V. acknowledges the financial assistance received through the Prime Minister's Research Fellowship (PMRF), awarded by the Ministry of Education, Government of India. R.M. extends appreciation to the Science and Engineering Research Board (SERB), Government of India, for funding this research under the Core Research Grant (CRG/2022/000663). T.M. acknowledges the support of the EURO HPC project (EHPC-DEV-2024D12-009). The authors also extend their appreciation to Skund Tewari and Sagnik Chatterjee for their insightful discussions, which contributed to the intellectual depth of this work.

\section*{Data Availability}
This paper has no additional data as it is a theoretical work.

\bibliographystyle{apsrev4-2-author-truncate}
\bibliography{ref}
\label{lastpage}

\end{document}